\documentstyle[multicol,aps,psfig]{revtex}
\begin{document}

\title{High-$p_T$ Hadron Spectra, Azimuthal Anisotropy and Back-to-Back 
Correlations \\
in High-energy Heavy-ion Collisions}
\author{Xin-Nian Wang}
\address{
Nuclear Science Division, MS70R0319,\\
Lawrence Berkeley National Laboratory, Berkeley, CA 94720}

\date{April 20, 2003}
\maketitle

\vspace{-1.5in}
{\hfill LBNL-52533}
\vspace{1.4in}

\begin{abstract}
The observed suppression of high-$p_T$ hadron spectra, finite azimuthal 
anisotropy, disappearance of jet-like back-to-back correlations, and 
their centrality dependence in $Au+Au$ collisions at RHIC
are shown to be quantitatively described by jet quenching within a
pQCD parton model. The difference between $h^{\pm}$ and $\pi^0$ 
suppression in intermediate $p_T$ is consistent with the observed
$(K+p)/\pi$ enhancement which should disappear at $p_T>6$ GeV/$c$.
The suppression of back-to-back correlations is shown 
to be directly related to the medium modification of jet 
fragmentation functions (FF) similar to direct-photon triggered FF's.

\end{abstract}

\begin{multicols}{2}

The degradation of high-$p_T$ partons during their propagation
in the dense medium can provide critical information necessary 
for detection and characterization of the strongly interacting 
matter produced in high-energy heavy-ion collisions. Because of 
radiative parton energy loss induced by multiple scattering,
the final high-$p_T$ hadron spectra from jet fragmentation are 
expected to be significantly suppressed \cite{wg92}.
Such a phenomenon, known as jet quenching, was observed for the 
first time in $Au+Au$ collisions at the Relativistic Heavy-ion 
Collider (RHIC) \cite{phenix-r,star-r}. One also observes 
the disappearance of back-to-back jet-like hadron 
correlations \cite{star-c} and finite azimuthal anisotropy \cite{star-v2} 
of high-$p_T$ hadron spectra. These three seemingly unrelated 
high-$p_T$ phenomena are all predicted as consequences of jet 
quenching \cite{wg92,mono,wangv2,gvw}. Together, they can provide 
unprecedented information on the properties of dense matter 
produced at RHIC. 

In this Letter, we will study these three high-$p_T$ phenomena
simultaneously within a lowest order (LO) pQCD parton model that 
includes initial nuclear $k_T$ broadening, parton 
shadowing and medium induced parton energy loss.
We point out that an enhanced $(K+p)/\pi$ ratio
leads naturally to different suppression of $h^{\pm}$ and $\pi^0$
spectra at intermediate $p_T$ range. We will also show that 
the suppression of back-to-back
correlations is directly related to the medium modification of 
hadron-triggered FF's similar to a direct-photon triggered FF \cite{whs}.

In a LO pQCD model \cite{wang98}, the inclusive high-$p_T$ hadron 
cross section in $A+A$ collisions is given by
\begin{eqnarray}
  \frac{d\sigma^h_{AB}}{dyd^2p_T}&=&K\sum_{abcd} 
  \int d^2b d^2r dx_a dx_b d^2k_{aT} d^2k_{bT} \nonumber \\
  & &  t_A(r)t_B(|{\bf b}-{\bf r}|) 
  g_A(k_{aT},r)  g_A(k_{bT},|{\bf b}-{\bf r}|) 
  \nonumber \\
  & & f_{a/A}(x_a,Q^2,r)f_{b/B}(x_b,Q^2,|{\bf b}-{\bf r}|) \nonumber \\
  & & \frac{D_{h/c}(z_c,Q^2,\Delta E_c)}{\pi z_c}  
  \frac{d\sigma}{d\hat{t}}(ab\rightarrow cd), \label{eq:nch_AA}
\end{eqnarray}
where $z_c=p_T/p_{Tc}$, $y=y_c$, $\sigma(ab\rightarrow cd)$ are 
parton scattering cross sections and $t_A(b)$ is the 
nuclear thickness function normalized to $\int d^2b t_A(b)=A$. 
We will use a hard-sphere model of nuclear distribution in this paper.
The $K\approx 1.3-2$ factor is used to account for higher order pQCD 
corrections.
The parton distributions per nucleon $f_{a/A}(x_a,Q^2,r)$
inside the nucleus are assumed to be factorizable into the parton 
distributions in a free nucleon given by the MRS D$-^{\prime}$  
parameterization \cite{mrs} and the impact-parameter dependent 
nuclear modification factor \cite{eks,lw02}. 
The initial transverse momentum
distribution $g_A(k_T,Q^2,b)$ is assumed to have a Gaussian form
with a width that includes both an intrinsic part in a nucleon and 
nuclear broadening. Details of this model 
and systematic data comparisons can be found in Ref.~\cite{wang98}.

As demonstrated in recent studies, a direct consequence of parton
energy loss is the medium modification of FF's \cite{guow00,sw02} 
which can be well approximated by \cite{ww02}
\begin{eqnarray}
D_{h/c}(z_c,Q^2,\Delta E_c) &=&(1-e^{-\langle \frac{\Delta L}{\lambda}\rangle})
\left[ \frac{z_c^\prime}{z_c} D^0_{h/c}(z_c^\prime,Q^2) \right.
 \nonumber \\
& &\hspace{-1.2in}
\left. +\langle \frac{\Delta L}{\lambda}\rangle
\frac{z_g^\prime}{z_c} D^0_{h/g}(z_g^\prime,Q^2)\right]
+ e^{-\langle\frac{\Delta L}{\lambda}\rangle} D^0_{h/c}(z_c,Q^2),
\label{modfrag} 
\end{eqnarray}
provided that the actual energy loss is about 1.6 times of
the input value. Here $z_c^\prime=p_T/(p_{Tc}-\Delta E_c)$,
$z_g^\prime=\langle \Delta L/\lambda\rangle p_T/\Delta E_c$
are the rescaled momentum fractions and $\Delta E_c$ is
the total parton energy loss during $\langle \Delta L/\lambda\rangle$
number of scatterings.
The FF's in free space $D^0_{h/c}(z_c,Q^2)$
are given by the BBK parameterization \cite{bkk}.

We assume a 1-dimensional expanding medium with a gluon 
density $\rho_g(\tau,r)$ that is proportional to the 
transverse profile of participant nucleons.
The average number of scatterings along the parton 
propagating path is then
\begin{equation}
\langle \Delta L/\lambda\rangle =\int_{\tau_0}^{\tau_0+\Delta L}
d\tau \sigma\rho_g(\tau,b,\vec{r}+\vec{n}\tau),
\end{equation}
where $\Delta L(b,\vec{r},\phi)$ is the distance a jet, produced at
$\vec{r}$, has to travel along $\vec{n}$ at an azimuthal 
angle $\phi$ relative to the reaction plane in a collision 
with impact-parameter $b$.

\begin{figure}
\centerline{\psfig{figure=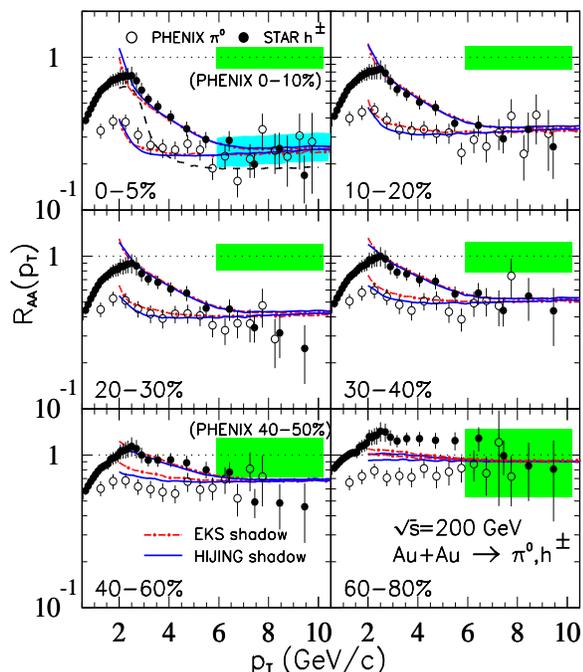,width=3.0in,height=3.5in}}
\caption{Nuclear modification factors for hadron spectra in $Au+Au$ collisions
as compared to data from STAR\protect\cite{star-r} and 
PHENIX \protect\cite{phenix-r}. See text for a detailed explanation.}
\label{fig1}
\end{figure}

According to recent theoretical studies \cite{gvw,sw02,ww02}
the total parton energy loss in a finite and expanding medium 
can be written as a path integral,
\begin{equation}
\Delta E\approx \langle \frac{dE}{dL}\rangle_{1d}
\int_{\tau_0}^{\tau_0+\Delta L} d\tau \frac{\tau-\tau_0}{\tau_0\rho_0}
\rho_g(\tau,b,\vec{r}+\vec{n}\tau),
\end{equation}
where $\rho_0$ is the averaged initial gluon density at $\tau_0$ 
in a central collision and $\langle dE/dL\rangle_{1d}$ 
is the average parton energy loss over a distance $R_A$
in a 1-d expanding medium with an initial uniform gluon 
density $\rho_0$. The corresponding energy loss 
in a static medium with a uniform gluon density 
$\rho_0$ over a distance $R_A$ is \cite{ww02}
$dE_0/dL=(R_A/2\tau_0)\langle dE/dL\rangle_{1d}$.
In the high-energy limit, the parton energy loss has a logarithmic
energy dependence \cite{loss1}. However, for a parton with 
finite initial energy the energy loss has a stronger energy 
dependence because of restricted phase space for 
bremsstrahlung \cite{glv00}. Detailed balance between induced 
gluon emission and absorption will further increase the energy 
dependence \cite{ww01} for a parton with finite initial energy. 
In this paper we will use an effective quark energy loss
\begin{equation}
 \langle\frac{dE}{dL}\rangle_{1d}=\epsilon_0 (E/\mu_0-1.6)^{1.2}
 /(7.5+E/\mu_0),
\label{eq:loss}
\end{equation}
from the numerical results in Ref.~\cite{ww01}.
The detailed balance reduces the effective parton energy loss and
at the same time increases the energy dependence. The threshold
is a consequence of gluon absorption that competes with
bremsstrahlung and effectively shuts off energy loss for
lower energy partons.

Shown in Fig.~\ref{fig1} are the calculated nuclear 
modification factors
$R_{AB}(p_T)=d\sigma^h_{AB}/\langle N_{\rm bin}^{AB}\rangle d\sigma^h_{pp}$
for hadron spectra ($|y|<0.5$) in $Au+Au$ collisions 
at $\sqrt{s}=200$ GeV, as compared to experimental data. Here,
$\langle N_{\rm bin}^{AB}\rangle=\int d^2bd^2r t_A(r)t_B(|\vec{b}-\vec{r}|)$.
To fit the observed $\pi^0$ suppression in the most 
central collisions, we have used (solid lines) $\mu_0=1.5$ GeV,
$\epsilon_0=1.07$ GeV/fm and $\lambda_0=1/(\sigma\rho_0)=0.3$ fm
with the new HIJING parameterization \cite{lw02} of parton shadowing.
The hatched area (also in other figures in this paper) indicates 
a variation of $\epsilon_0=\pm 0.3$ GeV/fm.
The hatched boxes around $R_{AB}=1$ represent experimental
errors in overall normalization. Alternatively, one has to
set $\mu_0=1.3$ GeV and $\epsilon_0=1.09$ when EKS
parameterization \cite{eks} of parton shadowing is 
used (dot-dashed lines).
Without parton energy loss, the spectra is slightly 
enhanced at $p_T=2-5$ GeV/$c$ due to
nuclear $k_T$ broadening even with parton shadowing. 

The flat $p_T$ dependence of the $\pi^0$ suppression is 
a consequence of the strong energy dependence of the
parton energy loss, which is also observed by other
recent studies \cite{postqm02}. The slight rise of $R_{AB}$
at $p_T<4$ GeV/$c$ in the calculation is due to the detailed
balance effect in the effective parton energy loss. In this
region, one expects the fragmentation picture to gradually 
lose its validity and other non-perturbative 
effects \cite{soft} become important that will give
an enhanced $(K+p)/\pi$ ratio in central $Au+Au$
collisions. To include this effect, 
we add a soft component to kaon and baryon FF's that
is proportional to the pion FF with a 
weight $\sim \langle N_{\rm bin}(b,r)\rangle/[1+\exp(2p_{Tc}-15)]$.
The functional form and parameters are adjusted so that
$(K+p)/\pi\approx 2$ at $p_T\sim 3$ GeV/$c$ in the most 
central $Au+Au$ collisions and approaches its $p+p$ value 
at $p_T>5$ GeV/$c$. The resultant suppression for
total charged hadrons and the centrality dependence 
agree well with the STAR data. One can relate $h^{\pm}$ 
and $\pi^0$ suppression via the $(K+p)/\pi$ ratio: 
$R_{AA}^{h^{\pm}}=R_{AA}^{\pi^0}[1+(K+p)/\pi]_{AA}/[1+(K+p)/\pi]_{pp}$.
It is clear from the data that $(K+p)/\pi$ becomes the same for
$Au+Au$ and $p+p$ collisions at $p_T>5$ GeV/$c$.
To demonstrate the sensitivity to the parameterized
parton energy loss in the intermediate $p_T$ region, 
we also show $R_{AA}^{h^{\pm}}$ in 0-5\% centrality (dashed line)
for $\mu_0=2.0$ GeV and $\epsilon_0=2.04$ GeV/fm without the 
soft component.

Since jets produced in the central core of the dense medium are 
suppressed due to parton energy loss, only those jets that are 
produced near the surface emerge from the medium.
The observed high-$p_T$ hadron multiplicity should 
be proportional to the number of surviving jets in the outer 
layer of the overlapped volume which in turn is approximately 
proportional to the total number of participant nucleons. 
The nuclear modification factor
normalized by $\langle N_{\rm binary}\rangle$ should 
decrease with centrality. This agrees well with the observed
centrality dependence. This surface emission picture also gives
a natural geometrical limit of the azimuthal anisotropy \cite{shuryak}.

\begin{figure}
\centerline{\psfig{figure=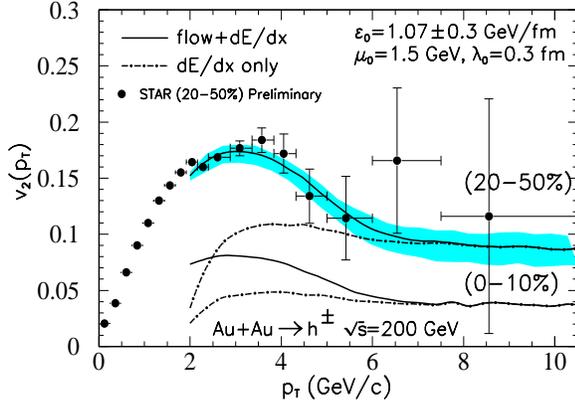,width=3.0in,height=2.1in}}
\caption{Azimuthal anisotropy in $Au+Au$ collisions
as compared to the STAR\protect\cite{snellings} 4-particle cumulant result.}
\label{fig2}
\end{figure}

In non-central collisions, the average path length of parton
propagation will vary with the azimuthal angle relative to
the reaction plane. This leads to an azimuthal
dependence of the total parton energy loss and therefore
azimuthal asymmetry of high-$p_T$ hadron spectra \cite{wangv2,gvw}.
Such asymmetry is another consequence of parton energy loss
and yet it is not sensitive to the nuclear $k_T$ broadening
and parton shadowing. Shown in Fig.~\ref{fig2}
is $v_2(p_T)$ (second Fourier coefficient of the
azimuthal angle distribution) of charged hadrons generated from
parton energy loss (dot-dashed) as compared to preliminary 
STAR data \cite{snellings}
using the 4-particle cumulant moments method \cite{ollitrault} which
is supposed to reduce non-geometrical effects such as
inherent two-particle correlations from di-jet production \cite{kovchegov}.
The energy loss extracted from high-$p_T$ hadron spectra 
suppression can also account for the observed azimuthal anisotropy at
large $p_T$. If the remaining $v_2$ at intermediate $p_T$
is made up by kaons and baryons from the soft component, 
one find that they must have $v_2\approx 0.23$ (0.11) for 20-50\% (0-10\%)
collisions. The total $v_2(p_T)$ is shown by the solid lines.

In the same LO pQCD parton model, one can also calculate di-hadron
spectra,
\begin{eqnarray}
  E_1E_2\frac{d\sigma^{h_1h_2}_{AB}}{d^3p_1d^3p_2}&=&\frac{K}{2}\sum_{abcd} 
  \int d^2b d^2r dx_a dx_b d^2k_{aT} d^2k_{bT} \nonumber \\
  & & \hspace{-1.0in} \int dz_cdz_d \,t_A(r)t_B(|{\bf b}-{\bf r}|) 
  g_A(k_{aT},r)  g_A(k_{bT},|{\bf b}-{\bf r}|) 
  \nonumber \\
  & & \hspace{-1.in} f_{a/A}(x_a,Q^2,r)f_{b/B}(x_b,Q^2,|{\bf b}-{\bf r}|)
  \nonumber \\
  & & \hspace{-1.0in} D_{h/c}(z_c,Q^2,\Delta E_c)
  D_{h/d}(z_d,Q^2,\Delta E_d) \nonumber \\
  & & \hspace{-1.0in} 
 \frac{\hat{s}}{2\pi z_c^2 z_d^2} \frac{d\sigma}{d\hat{t}}(ab\rightarrow cd)
 \delta^4(p_a+p_b-p_c-p_d),
 \label{eq:dih}
\end{eqnarray}
for two back-to-back hadrons from independent fragmentation
of the back-to-back jets. Set $p_{T1}=p_T^{\rm trig}$, we define 
a hadron-triggered FF as the back-to-back correlation with 
respect to the triggered hadron:
\begin{equation}
  D^{h_1h_2}(z_T,\phi,p^{\rm trig}_T)=
  p^{\rm trig}_T\frac{d\sigma^{h_1h_2}_{AA}/d^2p^{\rm trig}_T dp_Td\phi}
  {d\sigma^{h_1}_{AA}/d^2p^{\rm trig}_T},
\end{equation}
similarly to the direct-photon triggered FF \cite{whs} 
in $\gamma$-jet events. Here, $z_T=p_T/p^{\rm trig}_T$ and 
integration over $|y_{1,2}|<\Delta y$ is implied. 
In a simple parton model, the two jets should be
exactly back-to-back. The initial $k_T$ distribution
in our model will give rise to a Gaussian-like angular distribution.
In addition, we also take into account the intra-jet distribution
using a Gaussian form with a width of
$\langle k_T\rangle=0.8$ GeV/$c$.

\begin{figure}
\centerline{\psfig{figure=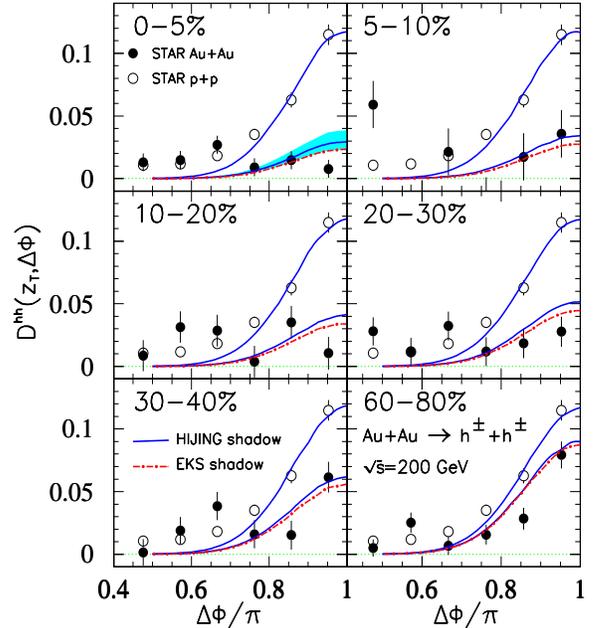,width=3.0in,height=3.3in}}
\caption{Back-to-back correlations for charged hadrons 
with $p^{\rm trig}_T>p_T>2$ GeV/$c$, 
$p^{\rm trig}_T=4-6$ GeV/$c$ and $|y|<0.7$ in $Au+Au$ (lower curves) 
and $p+p$ (upper curves)
collisions as compared to the STAR\protect\cite{star-c} data.}
\label{fig3}
\end{figure}

Shown in Fig.~\ref{fig3} are the calculated back-to-back correlations 
for charged hadrons in $Au+Au$ collisions as compared to the STAR 
data \cite{star-c}. The same energy loss that is used to calculate 
single hadron suppression and azimuthal anisotropy can also describe
well the observed away-side hadron suppression and its centrality
dependence. In the data, a background 
$B(p_T)[1+2v_2^2(p_T)\cos(2\Delta\phi)]$ from uncorrelated hadrons
and azimuthal anisotropy has been subtracted.
The value of $v_2(p_T)$ is measured independently while
$B(p_T)$ is determined by fitting the observed correlation in the
region $0.75<|\phi|<2.24$ rad \cite{star-c}.

Since most of jets produced in the central core 
of the colliding volume are suppressed, the triggered high-$p_T$ 
hadrons mainly come from jets produced near the surface 
traveling away from the dense core. These jets are not
quenched, as observed in STAR data \cite{star-c}.
Most away-side jets, however, will be suppressed as they go through
the dense core, except those that are propagating in directions 
tangent to the surface. This leads to a much smaller azimuthal
anisotropy of the away-side suppression. On the average,
both the magnitude of the away-side suppression 
and the centrality dependence should be similar to the single hadron 
suppression, as seen in the data.
Contrary to another study \cite{hirano}, we find that the
$p_T$ broadening associated with energy loss has no significant effect
on the observed back-to-back correlations, since those jets that have
large final-state broadening also have large energy loss and thus
are suppressed.

\begin{figure}
\centerline{\psfig{figure=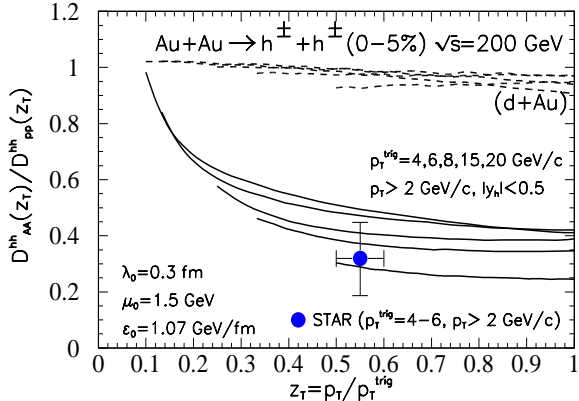,width=3.0in,height=2.1in}}
\caption{The suppression factor for hadron-triggered fragmentation
functions in central (0-5\%) $Au+Au$ (d+Au) collisions as compared to
the STAR data \protect\cite{star-c}.}
\label{fig4}
\end{figure}

Integrating over $\phi$, one obtains a hadron-triggered FF,
$D^{h_1h_2}(z_T,p_T^{\rm trig})=\int_{\pi/2}^{\pi} 
d\phi D^{h_1h_2}(z_T,\phi,p^{\rm trig}_T)$. Shown in Fig.~\ref{fig4} are
the suppression factors of the hadron-triggered FF's for different 
values of $p^{\rm trig}_T$ in central $Au+Au$ collisions
as compared to a STAR data point that is obtained by integrating
the observed correlation over $\pi/2<|\Delta\phi|<\pi$.
The dashed lines illustrate
the small suppression of back-to-back correlations
due to the initial nuclear $k_T$ broadening in $d+A$ collisions.
The strong QCD scale dependence 
on $p^{\rm trig}_T$ of FF's is mostly canceled in 
the suppression factor. The approximately universal shape
reflects the weak $p_T$ dependence of the hadron spectra 
suppression factor in Fig.~\ref{fig1}, due to a unique
energy dependence of parton energy loss.

In summary, we have studied simultaneously the suppression of
hadron spectra and back-to-back correlations, and high-$p_T$ azimuthal
anisotropy in high-energy heavy-ion collisions within a single LO pQCD
parton model incorporating current theoretical understanding
of parton energy loss. Experimental data of $Au+Au$ collisions
from RHIC can be quantitatively described by jet quenching in an
expanding medium. With HIJING (EKS) parton shadowing,
the extracted average energy loss for 
a 10 GeV quark in the expanding medium is 
$\langle dE/dL\rangle_{1d}\approx 0.85 (0.99) \pm  0.24$ GeV/fm, which
is equivalent to $dE_0/dL\approx 13.8 (16.1) \pm 3.9$ GeV/fm in a static and
uniform medium over a distance $R_A=6.5$ fm. This value 
is about a factor of 2 larger than a previous estimate \cite{ww02}
because of the variation of gluon density along the propagation
path and the more precise RHIC data considered here .

I would like to thank D. Hardtke, P. Jacobs and J. Klay
for helpful discussions. This work was supported by DOE under 
Contract No. DE-AC03-76SF00098.


\end{multicols}

\end{document}